\newcommand{\gev}{\, {\rm GeV}}
\newcommand{\tev}{\, {\rm TeV}}
\newcommand{\mev}{\, {\rm MeV}}

\newcommand{\kev}{\, {\rm keV}}
\newcommand{\beq}{\begin{equation}}
\newcommand{\eeq}{\end{equation}}
\newcommand{\bea}{\begin{eqnarray}}
\newcommand{\eea}{\end{eqnarray}}

\newcommand{\gsim}{\lower.7ex\hbox{$\;\stackrel{\textstyle>}{\sim}\;$}}
\newcommand{\lsim}{\lower.7ex\hbox{$\;\stackrel{\textstyle<}{\sim}\;$}}

\newcommand{\xpb}{\,{\rm pb}}

\documentclass[aps,prev,twocolumn,preprintnumbers,floatfix,nofootinbib]{revtex4}
\usepackage{graphicx}
\usepackage{mathrsfs}
\usepackage{amssymb}
\usepackage{verbatim}

\def\stacksymbols #1#2#3#4{\def\theguybelow{#2}
    \def\vp{\lower#3pt}
    \def\sp{\baselineskip0pt\lineskip#4pt}
    \mathrel{\mathpalette\intermediary#1}}

\def\intermediary#1#2{\vp\vbox{\sp
     \everycr={}\tabskip0pt
     \halign{$\mathsurround0pt#1\hfil##\hfil$\crcr#2\crcr
              \theguybelow\crcr}}}

\def\comment#1{}
\def\to{\rightarrow}

\def\u1x{U(1)_X}
\newcommand{\nc}{\newcommand}
\nc{\LL}{L}
\nc{\vv}{\tilde{v}}
\nc{\ccdot}{\!\cdot\!}
\nc{\gsm}{G_{SM}}
\nc{\vfive}{\mathbf{5}\oplus\mathbf{\overline{5}}}
\nc{\vten}{\mathbf{10}\oplus\mathbf{\overline{10}}}
\nc{\zhol}{Z^{\rm hol}}
\nc{\xfb}{\,{\rm fb}}

\begin{document}

%
%

\preprint{MCTP-06-11,MIFP-06-14}

\title{LHC and ILC probes of hidden-sector gauge bosons}

\author{Jason Kumar$^{a}$}
\author{James D. Wells$^{b}$}
\vspace{0.2cm}
\affiliation{
${}^a$ Physics Department, Texas A\&M University, College Station, TX 77843 \\
${}^b$ Michigan Center for Theoretical Physics (MCTP) \\
Department of Physics, University of Michigan, Ann Arbor, MI 48109}

\begin{abstract}
Intersecting D-brane theories motivate the existence of exotic $U(1)$ gauge bosons
that only interact with the Standard Model through kinetic mixing with
hypercharge.
We analyze an effective field theory description of this effect and describe the
implications of these exotic gauge bosons on precision electroweak, LHC and ILC
observables.

\end{abstract}

\maketitle


\maketitle


\setcounter{equation}{0}



\section{Exotic abelian symmetries}

There are many reasons to suspect that nature contains more abelian factors
than just hypercharge of the Standard Model (SM).  Traditionally, these
abelian groups were thought to arise as subgroups of larger unification
groups, such as $SO(10)\to G_{SM}\times U(1)$ or
$E_6\to G_{SM}\times U(1)^2$~\cite{Hewett:1988xc}.
This motivation
for extra $U(1)$ gauge groups led to studies of exotic $Z'$ bosons that coupled
at tree-level to the SM particles~\cite{Cvetic:1995rj,Leike:1997cw,Kang:2004bz}.

In this letter, we wish to emphasize the intersecting brane world motivation
for  extra $U(1)$ factors and
study their consequences for phenomenology within an effective theory framework.
In these constructions, one considers string theory compactified to four dimensions
with spacetime-filling D-branes wrapping cycles in the compact dimensions.  The
open strings which begin and end on the D-branes yield a low-energy gauge theory
which can potentially realize the Standard Model.  Although there are not yet
any known
intersecting brane models that have been completely worked out
and are free of phenomenological problems, this class
of string constructions is very broad, with a staggering number of potentially
viable vacua~\cite{Kumar:2006,Douglas:2006es}.  As such, it seems reasonable to assume
that there are models in this general class that can closely approximate
observed low-energy
physics, and a study of phenomenology generic to this class becomes quite interesting.

To begin with we note a basic fact:  a stack of $N$ coincident branes gives rise to the
gauge group $U(N)$, which is decomposed into $SU(N)\times U(1)$.  There are typically
many stacks of coincident branes in a complete, self-consistent string theory of
particle physics.  Unless the branes are at special points in the extra-dimensional space,
they will produce at least one abelian factor for every stack.  Of these, a few combinations
are anomaly free and massless (see, e.g.,~\cite{Marchesano:2004xz,Blumenhagen:2005mu,
Kumar:2004pv,Aldazabal:2000cn,Dijkstra:2004cc,Gmeiner:2005vz}).

Brane-world $U(1)$'s are special (compared to GUT $U(1)$'s) because there
is no generic expectation that SM particles will be charged under them.  When a
SM-like theory is constructed in brane-world scenarios, the SM particles generally
arise as open strings connecting one SM brane stack to another.
Exotic  non-SM branes usually carry the extra $U(1)$ factors.  However, there
are exotic states, called kinetic mesengers below, that can be charged under
the SM gauge group and the exotic $U(1)$.  These arise from open strings
connecting a SM brane stack to a hidden sector brane stack.  These states
can generate kinetic mixing between the $U(1)_Y$ and an exotic $U(1)$
symmetry that has phenomenological implications to be explored below.

Despite our D-brane motivations given above, we wish to transition
to an effective field theory description for our discussion of phenomenological
implications.  This, we believe, is a useful approach to string phenomenology:
identify a generic aspect of string theory (e.g., hidden-sector $U(1)$'s described
above), embed the specific idea into a more general
effective field theory framework, and then explore the phenomenological
implications of the wider range of parameters in
the effective theory.

In the next section we set out the effective theory description.   We then describe several of
the phenomenological implications of this straightforward but interesting
generic implication of D-brane scenarios.  The implications surveyed are
those of precision electroweak constaints, Large Hadron Collider (LHC)
detection prospects, and International Linear Collider (ILC) detection prospects.

\section{Effective theory description}

The framework described above gives rise to the possibility that an exotic $U(1)_X$
gauge symmetry exists that survives down to the TeV scale, but has no direct
couplings to SM particles.  Our effective field theory description
at a scale $\Lambda\gg m_Z$ has Standard Model (SM) gauge group
and an additional $U(1)_X$.  There are three sectors of matter
particles
\begin{itemize}
\item {\it Visible Sector:} Particles charged under the SM but not under $U(1)_X$.
The SM particles (quarks, leptons, Higgs, neutrinos) comprise this sector.
\item{\it Hidden Sector:} Particles charged under $U(1)_X$, but singlets
under the SM gauge groups.
\item{\it Hybrid Sector:} Particles charged under both the SM and $U(1)_X$
gauge groups, which we call {\it kinetic messengers} since they can induce
kinetic mixing between $U(1)_X$ and SM hypercharge.
\end{itemize}
For our purposes, we will assume that $U(1)_X$ is broken by a Higgs mechanism.
The mass-scale associated with $X$ breaking can be assumed for this discussion
to be tied to the
same mass scale that gives rise to electroweak symmetry breaking.
For example, softly broken supersymmetry masses could provide the requisite
Higgs masses for various sectors that all break the respective gauge symmetries
around the same supersymmetry breaking gravitino mass.

There are one-loop quantum corrections  that mix the kinetic terms
of $U(1)_Y$ and $U(1)_X$~\cite{Holdom:1985ag,delAguila:1995rb,Babu:1997tx,
Dienes:1996zr,Martin:1996kn,Rizzo:1998ut,Dobrescu:2004wz}.
We are then left with an effective Lagrangian at
scale $\mu$ for the kinetic terms of the form:
\beq
L_K=-\frac{1}{4}\hat B_{\mu\nu}\hat B^{\mu\nu}
-\frac{1}{4}\hat X_{\mu\nu}\hat X^{\mu\nu}
+\frac{\chi}{2}\hat B_{\mu\nu}\hat X^{\mu\nu}
\label{kinetic terms}
\eeq
where $\chi$ is given by
\beq
\chi=\frac{\hat g_Y\hat g_X}{16\pi^2} \sum_i Q^i_X Q^i_Y \log \left( \frac{m_i^2}{\mu^2}\right)
\eeq
and the sum $i$ is over all kinetic messenger states.

We cannot
say what value of $\chi$ is typical in the many possible brane-world
models of particle physics~\cite{Abel:2003ue}.  Although the above equation is a one-loop
expression, and perhaps expected to be small, the multiplicity of states
could be large enough to compensate for the one-loop suppression.  In a
different context, the issue of kinetic mixing among exotic $U(1)$'s was
investigated by Dienes, Kolda and March-Russell~\cite{Dienes:1996zr},
and it was estimated that $10^{-3}<\chi< 10^{-2}$; however,
this estimate may not be applicable for other approaches to model building.

One can choose a field redefinition $\hat X_{\mu},\hat Y_{\mu} \rightarrow
X_{\mu},Y_{\mu}$ that makes the
kinetic terms of eq.~\ref{kinetic terms}
diagonal and canonical.  The most
convenient choice of diagonalization is one
in which the couplings to $Y$ are independent of $Q_X$:
\beq
\left( \begin{array}{c}  X_\mu \\ Y_\mu \end{array}\right)
=\left( \begin{array}{cc} \sqrt{1-\chi^2} &  0 \\
-\chi & 1 \end{array}\right)
\left( \begin{array}{c} \hat X_\mu \\ \hat Y_\mu \end{array}\right).
\label{theta choice}
\eeq
The covariant derivative is now:
\bea
D_\mu \rightarrow \partial_\mu + i(g_XQ_X+\eta g_YQ_Y)X_\mu+ig_YQ_Y Y_\mu,
\eea
where
\beq
g_Y=\hat g_Y,~~g_X\equiv \frac{\hat g_X}{\sqrt{1-\chi^2}},~~\eta\equiv \frac{\chi}{\sqrt{1-\chi^2}}.
\eeq
(Note, $\eta\simeq \chi$ for small $\chi$.)
We are considering the case where $U(1)_X$ is broken
due to the veving of a hidden sector Higgs field $\Phi_X$ with
$Q_X\neq 0$ and $Q_Y=0$.  $X_\mu$ then gets a
mass $m_{X}^2\sim g_X^2\langle \Phi_X\rangle^2$, while $Y$ stays massless.
It is somewhat natural that $m_X$ is of order weak scale or TeV scale, especially
if the vev is controlled by supersymmetry breaking, as suggested earlier.
Note, the covariant derivative couples matter to $Y$ in the same
way as to $\hat Y$.  Thus, we can identify $Y$ as the hypercharge
gauge boson.

SM particles couple to $X^{\mu}$ with strength $\eta g_Y Q_Y$,
i.e. with couplings proportional to hypercharge.
This is an important phenomenological implication that enables
$X_\mu$ to be probed by experiments involving SM particles.
The $X_\mu$ behaves as a resonance of the hypercharge gauge
boson with somewhat smaller coupling; indeed, it may be confused
with an extra-dimension hypercharge gauge boson. It is also within
the general class of ``Y-sequential" gauge boson~\cite{Appelquist:2002mw}.

The $U(1)_X$ also couples to hidden sector fields at tree-level. However,
we assume the $U(1)_X$ gauge boson we are studying is too light to decay
into on-shell hidden sector particles or exotic kinetic messengers.
Intersecting brane models generally have multiple hidden sector gauge group
factors, which can break at different scales.  But light matter will appear
in chiral multiplets arising from strings stretching between different branes,
and their mass will be be set by the hidden-sector
gauge-symmetry breaking scales of the two gauge groups under which matter is charged.
If we study the hidden $U(1)_X$ which is broken at the lowest scale, the mass
of most hidden sector matter will be dominated by the higher symmetry breaking
scales of other gauge groups, and our assumption about the lightness of the
$U(1)_X$ gauge boson relative to other hidden matter is likely correct.
If this assumption is wrong,
the collider signatures that rely on branching fractions of $X$ boson
decays into SM particles would have to be adjusted.  Given the small
kinetic mixing angle we envision, if the $X$ does decay into long-lived hidden
sector states it is likely that the ILC searches described below for
$\gamma X$ production, where $\gamma$ recoils against ``nothing",
would be most useful.  Analogous LHC monojet or mono-photon signals
would need to be studied in that case as well.  If the $X$ decays into
long-lived charged, exotic messenger states, the quasi-stable massive
charged particle search strategies would be useful.

\section{Mass Eigenstates after Electroweak Symmetry Breaking}

When $SU(2)\times U(1)_Y$ breaks to $U(1)_{em}$, the $Z_\mu$ and
$X_\mu$ eigenvalues mix due to the small coupling of $X_\mu$ to
condensing Higgs
boson(s) that carry hypercharge.  The effects of this mixing
are minimal for the phenomenology of the $X_\mu$ boson at high-energy
colliders, except for two effects.  First, the mixing with the $Z$ boson
gives contributions to precision electroweak observables. Computing
observables from
effective Peskin-Takeuchi parameters~\cite{Peskin:1991sw,Holdom:1990xp,Babu:1997tx},
one finds the shifts
\bea
\Delta m_W & =  & (17\mev)\, \Upsilon \\
\Delta \Gamma_{l^+l^-} &  =  & -(8\kev)\, \Upsilon   \\
\Delta \sin^2\theta_W^{eff} &  = &  -(0.00033)\,\Upsilon
\eea
where
\beq
\Upsilon\equiv \left(\frac{\eta}{0.1}\right)^2\left(\frac{250\gev}{m_X}\right)^2.
\eeq
Experimental measurements~\cite{LEPEWWG}
of these most important electroweak observables
put limits on $|\Upsilon|\lsim 1$. Thus, for kinetic mixing of
$\eta \lsim {\cal O}(0.1)$
current precision electroweak observables do not constrain our effective theory
as long as $m_X$ is greater
than several hundred GeV. No meaningful bound for any
value of $\eta$ results if $m_X$ is
greater than about a TeV.
This fact is consistent with the precision electroweak analysis
of all other weakly coupled $Z'$ bosons that are summarized nicely
in the particle data group listings~\cite{PDG}.

The second consequence is that the mixing between the $Z$ and $X$ bosons
can change the hypercharge coupling of $X$ to SM particles.  This is
a subdominant effect for small $m_Z^2/m_{X}^2$, except it now allows
the $X$ mass eigenstate to decay into SM bosons.
After mixing, and assuming large $m_X$, one finds
\beq
\Gamma(WW)\simeq\Gamma(Zh)\simeq \eta^2(0.21\gev)\left(\frac{m_X}{1\tev}\right),
\eeq
each of which
is less than 2\% of the total width to fermions, calculated from
summing all
\beq
\Gamma(f\bar f)\simeq  N_c \eta^2(1.7\gev)(Y_{f_L}^2+Y_{f_R}^2)\left(\frac{m_X}{1\tev}\right).
\eeq
Because the branching fraction is not large,
we ignore bosonic decays in the subsequent analysis.

\section{LHC and Tevatron Probes}

We are now in a position to examine the possible collider signatures
of this scenario.  The process most
amenable to LHC analysis is on-resonance $pp \rightarrow X \rightarrow \mu \bar \mu$.
The predominant backgrounds are $pp \rightarrow \gamma^* /Z^*
\rightarrow \mu \bar \mu$.

For a hadron collider, observational bounds are somewhat
model-independent.  If we denote by $N_{X}$ the number of signal
events needed for a discovery signal, we find that the limit on
the mass of a discoverable $X$ is~\cite{Leike:1997cw,Kang:2004bz}
\bea
m_{X}^{lim} &\simeq& {\sqrt{s} \over A}
\ln \left({L \over s} {c_{X} C \over N_{X}}\right),
\eea
where the details of the
model are encoded in
\bea
c_{X} = {4\pi^2 \over 3}{\Gamma_{X}\over m_{X}}
B(\mu \bar \mu)
\left[B(u \bar u)
+{1\over C_{ud}} B(d \bar d)\right]. \nonumber
\eea
For a $pp$($p \bar p$) collider, $A=32(20)$, $C=600(300)$,
and in the kinematical
region of interest at LHC $C_{ud} \sim 2$ and $\sqrt{s}=14\tev$.
$L$ is the integrated luminosity.
If $m_{X} > m_{X}^{lim}$, $X_{\mu}$ cannot be observed at the
collider.
The logarithmic dependence of the detection bound
implies that this result is rather robust, somewhat
insensitive to variations in detector efficiency,
number of events needed for discovery, or small variations
in luminosity.

Substituting in the appropriate branching fractions $B(f \bar f)$
yields $c_{X}(X\rightarrow \mu \bar \mu) =
0.00456 \eta^2$ (for $m_X < 2m_{top}$, this will increase
by $< 15\%$).
We will fix an integrated luminosity of $100\xfb^{-1}$, which is
expected from LHC after a few years of high-luminosity running.
We then find
\bea
m_{X}^{\lim} &\simeq& 5.78\tev + (0.44\tev)
\ln \eta^2-(0.44\tev)\ln N_X \nonumber
\eea
at $100\xfb^{-1}$ integrated luminosity.
Equivalently, we may write the lower limit
on detectable kinetic mixing in terms of the
the mass of the $X$ and the number of signal
events $N_X$ as
\bea
\eta \geq \sqrt{N_X}e^{-6.61 +{m_{X}\over 0.88\tev}} \nonumber
\eea

To turn the above expressions into estimated bounds on $m_X$, we
need to determine how many signal events $N_X$ are needed to
discern the peak above background.  Since for much of the parameter
space, and in particular the parameter space near the edge of
detectability for small $\eta$, the $X$ boson is very narrow and we must
take into account experimental resolution.  The energy resolution of
an invariant $\mu^+\mu^-$ peak is expected to be no better than
a few percent~\cite{muon resolution}.  Thus, we cannot choose bin
sizes too small to maximize signal events over background events.
For our parameter space, a minimum bin size of $50\gev$ will become
appropriate for any $m_X\lsim 2\tev$, which will be about the maximum
value of $m_X$ detectable if $\eta\lsim 0.1$.  The muon resolution decreases
as $m_X$ decreases, but the electron resolution gets better. Thus, we could
substitute $e^+e^-$ decay analysis for very massive $m_X$ all the way up to
$\eta\sim 1$ and $m_X\lsim 5\tev$, which is approximately
the maximum value of $m_X$ that one could hope for detecting a weakly
coupled $Z'$ boson at the LHC~\cite{Dittmar:2003ir}.  As we are interested
in probing the smallest values of $\eta$ it will not be necessary to consider
that possibility further.

\begin{figure}
\includegraphics[width=6.0cm]{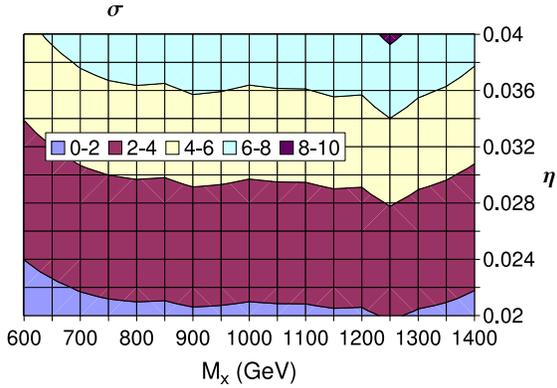}
\caption{LHC detection prospects for $100\xfb^{-1}$ of integrated luminosity
in the $\eta$-$M_X$ plane.
The countours are of signal significance, which exceeds $5$ only when
$\eta\gsim 0.03$.}
\label{figure1}
\end{figure}

Using Pythia~\cite{Pythia} to simulate the SM background in 50 GeV bins,
we can then plot ${N_X\over \sqrt{N_{bgd.}}}$ at LHC as a function
of $\eta$ and $m_{X}$(Fig.\ref{figure1}).
Detection at the LHC requires $N_X/\sqrt{N_{bgd}}>5$ in a single
bin normalized to a smooth SM background distribution.
We see that realistic demands for a signal which can
be distinguished from the background will require
$\eta > 0.03$.  Below the top threshold, the bounds
on $\eta$ may shift by $<8\%$.
Although this is unfortunately not
a good probe when compared to naive one-loop
perturbative estimates, the multiplicity of
kinetic messengers may enable $\eta>0.03$ and so
should be studied with care at the LHC.

The analysis at Tevatron is similar, but with different
parameters (accounting for differences in specifications
and for a $p\bar p$ collider).  We now have $C_{ud}=25$
and $\sqrt{s}=2\tev$. Assuming an integrated luminosity
of $8\xfb^{-1}$, the $\eta$ sensitivity is
\bea
\eta_{teva.} \geq \sqrt{N_X}e^{-6.7 +{m_{X}\over 0.2\tev}} \nonumber
\eea
For detection we demand at least a 5$\sigma$ signal above
background, or at least 10 events above background (whichever is
larger).
At $m_X=500\gev$ detection could only occur if $\eta>0.07$
for this high luminosity. As $m_X$ increases, the sensitivity
limits on $\eta$ degrade rapidly.

\section{ILC Prospects}

Given the challenge for LHC detection posed by
small kinetic mixing, one might hope
that ILC can do better.  An $e^+ e^-$ collider will
generally trade away $\sqrt{s}$ for higher luminosity
($\sim 500\xfb^{-1}$) and a
cleaner signal.  One does not produce an $X$ on resonance, of
course, unless its mass is less than the center of mass
energy, which we assume here to be $500\gev$.

The basic process we are interested in is
$e^-e^+ \rightarrow \mu \bar \mu$ through $\gamma^*/Z^*/X^*$.
The observable that provides perhaps the most useful signal
in this case is the total cross-section\cite{Leike:1993ky}(the
forward-backward asymmetry and left-right polarization do
not appear to provide qualitative improvement).
We may write the total cross-section as
\bea
\sigma_{\rm tot}(f\bar f) & = & \frac{N_c}{48\pi s}\sum_{n,m}
\frac{g^2_n g^{*2}_m s^2I^{f}_{m,n}}{(s-m^2_{V_n})(s-m^2_{V_m})^*}
\eea
where
\bea
I^{f}_{m,n}& = &
(L^e_n L^{e*}_m+ R^e_n R^{e*}_m)(L^f_n L^{f*}_m + R^f_n R^{f*}_m)
\eea
and the coupling of the $V_n$ boson to the fermions $f\bar f$ is
given by $ig_n\gamma^\mu(L^f_n P_L+R^f_nP_R)$.
If $m_X > 500\gev$, this observable will provide the
dominant signal.
Near the resonance, the signal is enhanced and we should replace:
\beq
\frac{1}{s-m^2_V}\Longrightarrow
\frac{(s-m^2_V)-i\Gamma_V\sqrt{s}}{(s-m^2_V)^2+s\,\Gamma_V^2} \Longrightarrow
\frac{-i}{\Gamma_V m_V}.
\eeq

\begin{figure}
\includegraphics[width=6.0cm]{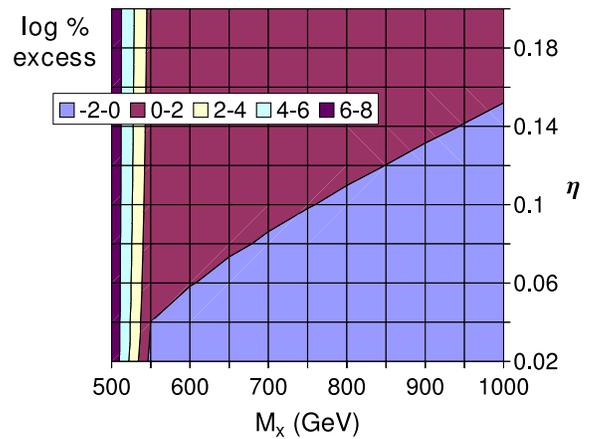}
\caption{Deviations of $e^+e^-\to \mu\bar \mu$
at ILC at $\sqrt{s}=500\gev$ for $500\xfb^{-1}$ integrated luminosity
are represented in this plot as contours of the $\log_{10}(\%)$ of the
excess of events produced compared to SM expectations.  The line along the interface
of the blue and maroon regions represents a $10^0=1\%$ (or $\sim 5\sigma$) deviation.}
\label{figure2}
\end{figure}

Our strategy is to compare the inclusive cross-section for $X$ production
to the pure Standard Model background.  Our
criterion for a signal
detection is at least 1\% deviation from SM expectations, in order
not to run afoul of systematic uncertainties.  Recall, we are
assuming $500\xfb^{-1}$ of integrated luminosity, and so the corresponding
statistical significance of the signal is
$\sim {\cal L} \sigma_S/\sqrt{{\cal L}\sigma_B}=0.01\sqrt{{\cal L}\sigma_B}\simeq 4.7
\sim5\sigma$, given the SM cross-section $\sigma_{tot}=447$fb~\cite{COMPHEP}.
Fig.~\ref{figure2} shows the deviations of $e^+e^-\to \mu\bar \mu$
at ILC at $\sqrt{s}=500\gev$ for $500\xfb^{-1}$ integrated luminosity.
Increasing values of $M_X$ can be probed only by increasing values of the
mixing parameter $\eta$.  For example, $M_X=750\gev$ ($1000\gev$) can be
probed for values of $\eta$ as low as $0.10$ ($0.15$).

If $m_X<500\gev$, then
we should instead consider the hard-scattering process
$e^-e^+ \rightarrow \gamma X \rightarrow \gamma f\bar f$.
The emission of a hard photon will allow us to scatter through a
resonance of the $X$, enhancing the cross-section and yielding a
cleaner signal.  This is a leading
order calculation, as radiation of more photons would serve to enhance
both the signal and backgrounds we calculate for the single photon case.

\begin{figure}[t]
\includegraphics[width=6.0cm]{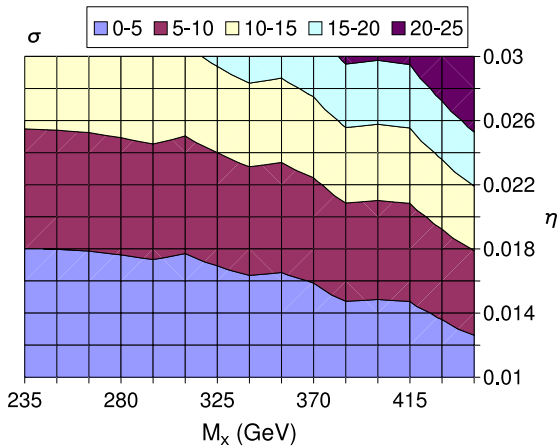}
\caption{Signal significance plot of $e^+e^-\to \gamma X\to \gamma\mu \bar \mu$
at ILC $\sqrt{s}=500\gev$ with $500\xfb^{-1}$ integrated
luminosity. We assume that the $m_{\mu\bar\mu}$ can be measured to within 2\%. }
\label{figure3}
\end{figure}

The differential cross-section of $\gamma X$ production is
\bea
{d\sigma \over dx} &=& {\alpha (c_L ^2 +c_R ^2)[ (s+m_X^2)^2 + (s-m_X^2)^2 x^2]\over
4s^2(s-m_X ^2)(1-x^2)}
\eea
where $x\equiv \cos\theta$ and $i\gamma^\mu (c_{L}P_L+c_R P_R)$
are the couplings of the left and right handed
electrons to $X_\mu$.  We choose a standard $|\cos \theta|<0.95$
angular cut.
The signal we analyze~\cite{signal comment} is $X \rightarrow \mu \bar \mu$,
so we must multiply by the appropriate branching fraction,
$B(\mu \bar \mu)=0.12$.
Substituting in the couplings we find
\bea
\sigma(\gamma X) &=& \eta^2 \, \frac{1.26\times 10^{-3}}{s^2 (s-m_X^2)}
F(s,m_X,x_0), ~~{\rm where} \nonumber\\
F(s,m_X,x_0) &= & (s^2 + m_X ^4)\tanh^{-1} x_0 -{x_0\over 2}(s-m_X ^2)^2\nonumber
\eea
and $x_0=|\cos \theta_{min}|=0.95$.
Comparing this signal to the Standard Model background
$e^+ e^- \rightarrow \gamma \mu \bar \mu$ gives us signal
significance at ILC for $m_X < 500\gev$ as well(Fig.\ref{figure3}).
We find that kinetimatically accessible $X$ bosons at ILC
can be probed and studied perhaps better than at the LHC.
Having the ILC data, along with the LHC data, can significantly
help us understand all the properties of an exotic massive weakly
coupled vector boson~\cite{Freitas:2004hq}.

An analysis of LEP data proceeds in a similar manner.  For $m_X >\sqrt{s}$,
detection from LEP data is highly disfavored.  For smaller $m_X$, however,
resonance production (through hard photon emission) is allowed, which
favors detection at LEP over hadron colliders such as LHC or the Tevatron.
(We assume $725\xpb^{-1}$ at $\sqrt{s}=206\gev$.) But
even at these low $m_X$ values ILC would provide better detection sensitivity.

\section{Outlook}

\begin{figure}[t]
\includegraphics[width=6.0cm]{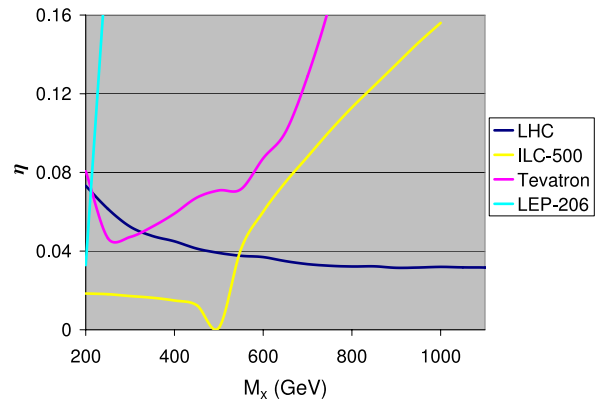}
\caption{Detection plot of estimated $5\sigma$ confidence level
of $X$-boson that kinetically mixes with hypercharge.
Detection for Tevatron ($8\xfb^{-1}$),
LHC ($100\xfb^{-1}$), LEP ($\sqrt{s}=206\gev$ and $725\xpb^{-1}$),
and ILC ($\sqrt{s}=500\gev$ and
$500\xfb^{-1}$) can occur at points above their respective lines.}
\label{figure4}
\end{figure}

We can summarize our results with a detection plot (Fig.\ref{figure4}).
We see that for $m_X \gsim 550\gev$, LHC is more capable of detecting
$X_{\mu}$ in our scenario,
while for $m_X \lsim 550\gev$ ILC-500 is more sensitive.  Detection within
the regime favored by a naive perturbative estimate of kinetic mixing
($\eta \sim 10^{-3}-10^{-2}$)
appears difficult at the LHC, but perhaps possible at the ILC as
long as the gauge boson mass is at or below the center of mass energy
of the ILC. Of course, a higher energy ILC (1 TeV and higher) will help
the search for higher-mass $X$ bosons.
As we emphasized earlier, however, the amount of kinetic mixing
can vary dramatically from one model to the next, depending on the
multiplicity of the kinetic messengers, and thus all regions of
phase space are candidates for discovery.

{\it Acknowledgments.}
We gratefully acknowledge S.Abel, B.Dutta, B.Holdom, P.Langacker,
D.Morrissey, A.Rajaraman, T.Rizzo, G.Shiu and M.Toharia for helpful
discussions.
This work is supported by the Department of Energy, NSF
Grant PHY-0314712 and the MCTP.



\begin{thebibliography}{99}
\bibitem{signal comment}
 We choose $\mu^+\mu^-$ as a specific
final state, but detection of the $X$ boson can come from any decay
channel (including invisible) by looking for the peak in photon-recoil
spectrum.  Definitive final states, such as $\mu^+\mu^-$ invariant mass
peak however are sharp signals of the $X$-boson mass and branching
fraction into $\mu^+\mu^-$.
\end{thebibliography}
\end{document}